\documentclass[fleqn,usenatbib]{mnras}

\usepackage{newtxtext,newtxmath}

\usepackage[T1]{fontenc}

\DeclareRobustCommand{\VAN}[3]{#2}
\let\VANthebibliography\thebibliography
\def\thebibliography{\DeclareRobustCommand{\VAN}[3]{##3}\VANthebibliography}

\usepackage{graphicx}	
\usepackage{amsmath}	
\usepackage{siunitx}	
\usepackage{physics}	

\usepackage{csquotes}
\usepackage{ulem}
\usepackage{xcolor}

\title[II lab tests for IACT measurements]{Optical intensity interferometry lab tests in preparation of stellar diameter measurements at IACTs at GHz photon rates}

\author[A. Zmija et al.]{
Andreas Zmija,$^{1}$\thanks{E-mail: andi.zmija@fau.de}
Naomi Vogel,$^{1}$
Gisela Anton,$^{1}$
Dmitry Malyshev,$^{1}$
Thilo Michel,$^{1}$
\newauthor
Adrian Zink,$^{1}$
Stefan Funk,$^{1}$
\\
$^{1}$Erlangen Centre for Astroparticle Physics, Friedrich-Alexander-Universität Erlangen-Nürnberg, Erwin-Rommel-Str. 1, Erlangen D 91058, Germany\\
}

\date{This is a pre-copyedited, author-produced PDF of an article accepted for publication in Monthly Notices of the Royal Astronomical Society following peer review. The version of record [\cite{10.1093/mnras/stab3058}] is available online at: https://academic.oup.com/mnras/article/509/3/3113/6408477.}

\pubyear{2021}

\begin{document}
\label{firstpage}
\pagerange{\pageref{firstpage}--\pageref{lastpage}}
\maketitle

\begin{abstract}
Astronomical intensity interferometry enables quantitative measurements of the source geometry by measuring the photon fluxes in individual telescopes and correlating them, rather than correlating the electromagnetic waves' amplitudes. This simplifies realization of large telescope baselines and high angular resolutions. Imaging Atmospheric Cherenkov Telescopes (IACTs), intended to detect the optical emission of $\gamma$-ray induced air showers, are excellent candidates to perform intensity correlations in the optical at reasonable signal-to-noise ratios. The detected coherence time is on the scale of $10^{-12}$ to $10^{-15}$~seconds - depending on the optical bandwidth of the measurement - which challenges the detection system to work in a stable and accurate way. We developed an intensity interferometry setup applicable to IACTs, which measures the photo currents from photomultipliers and correlates them offline, and as such is designed to handle the very large photon rates provided by the telescopes. We present measurements in the lab simulating starlight using a xenon lamp and measured at different degrees of temporal and spatial coherence. Necessary calibration procedures are described with the goal of understanding the measurements quantitatively. Measured coherence times between $5\,$femtoseconds (corresponding signal-to-background ratio $5\cdot10^{-7}$) and $110\,$femtoseconds (signal-to-background ratio $10^{-5}$) are in good agreement with expectations, and so are the noise levels in the correlations, reaching down to $6 \cdot 10^{-8}$, after measurements between $30\,$minutes and $1\,$ hour.
\end{abstract}




\section{Introduction}
In high resolution astrophysics, multiple telescopes are combined to form an interferometric array in order to overcome the diffraction limit of a single telescope \citep{Monnier_2003}. Conventionally, the technique of amplitude interferometry is exploited to observe the first order correlation pattern of the astronomical light source. Prominent examples in the optical/infrared regime are VLTI or CHARA, which provide baselines between telescopes of up to $331\,$m \citep{CHARA_description}. These observations are limited in the baseline between the telescopes and therefore in optical resolution due to atmospheric fluctuations and the difficulty of combining the light paths of the different telescopes at the necessary sub-wavelength precision.

A possible improvement in angular resolution for bright stars/astronomical objects is the technique of intensity interferometry, where light intensities - number of photons - in the different telescopes are recorded and second order correlations are observed \citep{HBT1956}. Since the necessary precision of controlling the different light path lengths is related to the distance light travels within the system's time resolution, which for time resolutions of some nanoseconds translates to $\approx\,$meters, above mentioned difficulties are strongly alleviated, allowing for kilometer baseline interferometry in the optical band \citep{Dravins_kilometer_baseline}.

Such intensity correlations were carried out first by Hanbury Brown and Twiss with the Narrabri Stellar Intensity Interferometer in the 1960s and early 1970s \citep{brown1967stellar, HBT_32}. The main challenge in intensity interferometry is detecting the small correlation signal with a reasonable significance ideally employing good time resolutions of the photon signal detection and processing electronics as well as large telescopes. Since none of these existed at the time, this technique wasn't pursued further for decades \citep{bojer2021quantitative}.

Due to tremendous improvements in technology \citep{DRAVINS2013331}, new interest in this topic has evolved and research groups have formed over the last years. In 2017, temporal correlation signals of three stars have been successfully measured with the $1\,$m diameter telescopes at the C2PU observatory at the Calern plateau at nanosecond time resolution \citep{rivet2018optical}. Furthermore, with AQUEYE+ and IQUEYE astronomical instruments attachable to optical telescopes are developed for sub-nanosecond timing resolution observations, which are also suited for stellar intensity interferometry measurements \citep{aqueye_ii}.

Aiming for high S/N measurements at short measurement times, the use of Cherenkov telescopes for intensity interferometry is promising \citep{Nu_ez_2010}. Even though they often cannot compete in terms of timing resolution with smaller optical telescope instruments, they have the advantage of providing huge light collection areas, enabling measurements at high photon count rates and thus decreasing the necessary measurement time - provided that systematic effects can be controlled.

In 2019, intensity correlation measurements at the MAGIC Cherenkov telescopes revealed significant photon bunching peaks of different stars after only a few minutes of effective measurement time \citep{magic_ii}. In the same year, observations at the VERITAS Cherenkov telescopes demonstrated the power of such arrays by measuring the photon correlations of two stars, which have also been measured by \cite{HBT_32}, at many different projected telescope baselines. Thus the angular diameters of these stars were determined with increased precision compared to the original Hanbury Brown-Twiss measurements, but more than 10 times faster \citep{veritas_II}.

In preparation for a measurement campaign at such telescope arrays, we present laboratory tests of an intensity interferometer which can be operated at the expected high photon rates of up to GHz.

As light source a xenon arc lamp is used providing a wide spectrum and high color temperatures simulating the star. Two different pinhole sizes in front of the lamp create different spatial coherences - simulating two different telescope baselines.

Correlation measurements are carried out using a beam splitter between the two photo-detectors and a $2\,$nm FWHM optical filter. This filter width is an intermediate approach. Smaller optical filters typically are used in non-astronomical intensity interferometry as well as in the previously mentioned stellar intensity interferometry measurements with smaller optical telescopes, which results in an increased temporal coherence of the light. However, since optics need to be precisely adjusted for narrow optical filters, which is very challenging at IACTs, $2\,$nm is rather narrow (but still reasonable) compared to the filter widths used at the MAGIC ($36\,$nm) and VERITAS ($13\,$nm) telescopes \citep{magic_ii, veritas_II}.
An additional measurement with a $36\,$nm filter, the same as used in MAGIC, has also been performed to test the setup at very high photon rates but small temporal coherence.

\section{Observables}

The observable in astronomical intensity interferometry is the second-order correlation function

\begin{equation}\label{eq:g2}
    g^{(2)}\left( \mathbf{r_0} , \mathbf{r_1}, \tau \right) = \frac{ \langle I\left (\mathbf{r_0}, t \right) I \left( \mathbf{r_1}, t+\tau \right) \rangle}{ \langle I\left (\mathbf{r_0}, t \right) \rangle \langle I \left( \mathbf{r_1}, t \right) \rangle} = 1 + g_s^{(2)}(\mathbf{r_1}-\mathbf{r_0})\cdot g_t^{(2)}(\tau)
\end{equation}
where the intensities $I(\mathbf{r_i})$ at telescope positions $\mathbf{r_i}$ are measured and correlated in the time averaged product for a given time difference $\tau$. $g_s^{(2)}$ and $g_t^{(2)}$ are the spatial and temporal parts of the correlation function. 
The temporal correlation function is connected to the spectrum of the observed photons via Fourier transform \citep{loudon2000quantum}. When spatial coherence conditions are satisfied, an excess of photon correlations at small time differences ($|\tau| \lessapprox \tau_c \approx 1/\Delta\nu$) is measured compared to the number of random photon correlations at larger time differences, where $\tau_c$ is called coherence time and $\Delta\nu$ is the optical bandwidth of observed photons. When applying nanometer-bandwidth optical filters, the coherence time is on the order of $\tau_c < 1\,$ps, demonstrating the need of fast electronics for detecting the correlation signal. The measured coherence time can be defined as the integral of $g_t^{(2)}$:

\begin{equation}\label{eq:coh_time}
    \tau_c := \int_{-\infty}^{+\infty} g_t^{(2)}(\tau) \,d\tau = \int_{-\infty}^{+\infty} \left(g^{(2)}(\mathbf{r_0}=\mathbf{r_1},\tau) - 1\right) \,d\tau
\end{equation}
According to the Van Cittert-Zernike theorem the spatial part of the correlation function is related via Fourier transform to the emission profile of the source \citep{mandel1995optical}, resulting in high coherence for small detector separations and angular source diameters. Changing the detector separation, meaning the telescope baseline projected on the observation direction, allows for sampling the spatial correlation function and drawing back to e.g. the angular size of the source. While it is not possible for us in the laboratory to realize baseline separation-induced spatial coherence losses, which is desirable in order to test the behaviour of the interferometer, we instead restrict the size of the light source by adding pinholes of different diameters. This results in different degrees of spatial coherence since the source extent is already (partially) resolved within the size of the detectors.

The fluctuations of $g^{(2)} (\tau\gg\tau_c)$ are due to random photon correlations and can be quantified using the root mean square value of the $g^{(2)}$ datapoints assuming Poissonian shot noise statistics

\begin{equation}\label{eq:RMS_timetagging}
    \sigma_{g^{(2)}} = 1/\sqrt{R_0R_1\Delta tT}
\end{equation}
where $R_i$ are the photon rates at detectors $i$, $\Delta t$ is the time bin width in the measurement and $T$ is the integrated measurement time. We call this fluctuation baseline fluctuation (of the $g^{(2)}$ function).

We have previously shown that we are able to quantitatively detect such a correlation signal with an intensity interferometer in the lab measuring photon rates of about $10\,$MHz per detector at measurement times of $40\,$hours \citep{Zmija:20}. These measurements were performed using photon time tagging electronics, which enabled correlations of the arrival times of photons.

Using the collection area of Cherenkov telescopes, like the High Energy Stereoscopic System (H.E.S.S.), which is on the order of $100\,$m$^2$ \citep{bernlohr2003optical}, or the future Cherenkov Telescope Array (CTA), photon rates beyond hundreds of MHz are expected from bright stars strongly decreasing necessary measurement times to a reasonable time range of a few minutes.

Using photometry calibration data given by \cite{1979PASP...91..589B}, for a $2\,$nm filter centered at $465\,$nm ($B\,$-band) and a single $100\,$m$^2$ telescope, the expected rates at each photodetector after the light is split by a beam splitter can be estimated as

\begin{equation}
    R = \eta \cdot 13.83 \cdot 100^{-m_\textnormal{V}/5}\,\textnormal{GHz}
\end{equation}
where $m_\textnormal{V}$ is the star's apparent magnitude and $\eta$ is the total efficiency, which consists of the atmospheric transmission, telescope and setup transmission and detector quantum efficiency. Estimating $\eta = 0.1$, the expected photon rate of a magnitude zero star is $1.38\,$GHz per detector, for a magnitude $2$ star it is $220\,$MHz. 
At these rates, single photon time tagging is technically not possible. Instead, photon currents are recorded and correlated. 

In order to prepare for a measurement campaign with such telescopes, lab experiments with a new interferometer have been carried out at up to GHz photon rates using photon current correlations in order to determine the temporal correlation function.

\section{Measurement setup}
Correlation measurements are done by recording the digitized photomultiplier (PMT) currents and cross-correlating them offline for different time differences $\tau$ to obtain the temporal second order correlation function $g^{(2)}(\tau)$.

\subsection{Optical setup}

\begin{figure}
\centering
\includegraphics[width=0.8\linewidth]{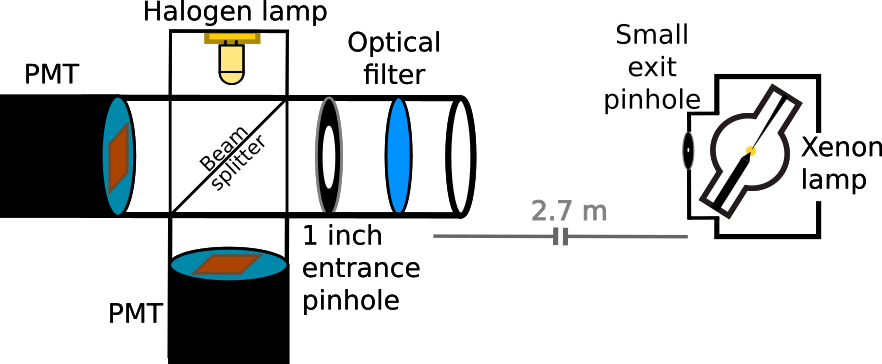}
\caption{Schematic description of the setup.}
\label{fig:interferometer}
\end{figure}

Fig.~\ref{fig:interferometer} shows the setup of the interferometer used for the lab measurements. Thermal light of the xenon lamp (XBO 75 W/2 OFR) passes a small exit pinhole, which determines the amount of optical coherence of photons arriving at the PMTs. Two exit pinholes of sizes $30\,$\textmu m and $75\,$\textmu m diameter are used to create spatial coherence of the light of $g_s^{(2)} = 0.8$ and $g_s^{(2)} = 0.36$ respectively at the interferometer.

The interferometer part consists of a lightproof stackable $2\,$inch diameter tube system to prevent straylight from entering the setup. An optical filter is placed in front of the $1\,$inch entry pinhole at the beam splitter. An Alluxa 465-2 OD4 \citep{alluxa} interference filter ($2\,$nm FWHM) as well as a Semrock FF01-432/36-25 ($36\,$nm FWHM) \citep{Semrock} are used in the measurements defining the detected wavelength spectra. The light is then split by a non-polarizing beam splitter cube (Thorlabs BS031) and directed at 90$^\circ$ angle to the two PMTs.

On the remaining side of the beam splitter a halogen lamp is placed. It is usually switched off during correlation measurements, but is used for calibrating the setup, as described in section \ref{sec:calibration}.

\subsection{Photo detection}

\begin{figure}
\centering
\includegraphics[width=1\linewidth]{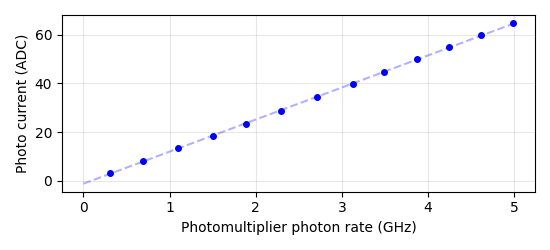}
\caption{Gain stability of a used photomultiplier}
\label{fig:gain}
\end{figure}

We use photomultipliers of Type Hamamatsu R11265U-300, which have the shape of a square with $23\,$mm side length and $>30$ per cent quantum efficiency at the measured wavelengths \citep{hamamatsu}. The PMT signals are amplified by a factor of $10$ before being digitized. A linear response of the photomultiplier currents with increasing photon rate is important for calibrating the interferometer. To prevent gain drop at high photon fluxes the last four dynodes of the PMTs are supplied with a stabilizing voltage.
The behaviour was tested by illuminating a PMT with an LED at varying power. The PMT was installed on one side of the beam splitter and a calibrated photo-diode on the other side, whose current was monitored. At low photon rates and high PMT gain ($900\,$V supply voltage), where single photon pulse counting in the digitized PMT data is possible, a calibration between the diode current and the PMT rate was performed in order to be able to calculate the PMT photon rates in Fig.~\ref{fig:gain} from the diode current. Afterwards the PMT gain was decreased to $680\,$V, which allows for GHz rates without causing damage to the PMT. 
Fig.~\ref{fig:gain} shows the obtained result demonstrating the gain stability for photon rates of up to $5\,$GHz.

\subsection{Digitizing electronics}

We used a M4i.2212-x8 card from Spectrum instrumentation as digitizer. It consists of four input channels, two of which were used for recording the two PMT signals. It allows for 8 bit sampling with a sampling speed of $1.25$ GS$\,$s$^{-1}$ - $0.8\,$ns time bin width respectively \citep{spectrum_instrumentation}. However, for data reduction and increased correlation speed, only $625\,$MS$\,$s$^{-1}$ - $1.6\,$ns - sampling was used in the measurements. In view of the use in H.E.S.S. \citep{bernlohr2003optical} or CTA MST \citep{schlenstedt2014medium} telescopes, which have optical time spreads of the mirrors between $\approx$ one and a few ns, the reduction of sampling speed is appropriate. For the 8 bit sampling a voltage range of $\pm\,200\,$mV was used.

\section{Analysis chain and measurement procedure}
\subsection{Raw data correlation}

\begin{figure}
\centering
\includegraphics[width=1\linewidth]{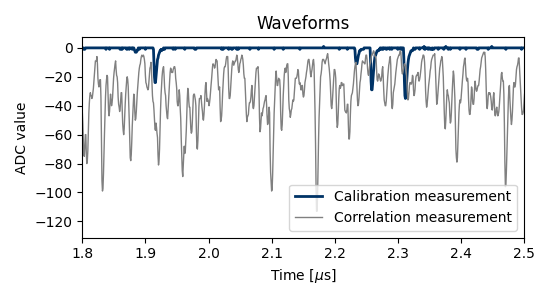}
\caption{Waveforms of a low-rate (blue calibration) and a high-rate (grey correlation) measurement}
\label{fig:waveforms}
\end{figure}

Fig.~\ref{fig:waveforms} shows two waveforms, which describe the digitized photon current in one PMT channel, in an arbitrary time window. The waveform in blue was taken at a low photon rate and illustrates the response of the photomultiplier to single photons. However, in the correlation measurements, photon rates of several $100\,$MHz were measured resulting in a waveform where many single photon pulses overlap, as seen in Fig.~\ref{fig:waveforms} (grey).

The correlation algorithm follows the conventional (at first step unnormalized) signal cross correlation. For time difference $\tau=0$, the data vectors $\mathbf{A}(t)$ and $\mathbf{B}(t)$ of the waveforms of the two PMTs are multiplied, while for $\tau\neq 0$ one channel vector is shifted by $\tau$. The output is an unnormalized spectrum of photon coincidences at each time difference, which we refer to as $G^{(2)}(\tau)$.

\begin{equation}\label{eq:G2}
    G^{(2)} (\tau) = \mathbf{A} (t) \boldsymbol{\cdot} \mathbf{B} (t+\tau) = \sum_{i=1}^{N} A(t_i)B(t_i+\tau)
\end{equation}
It already enables signal-to-noise investigations, but for obtaining physical parameters, such as the coherence time, a normalization has to be applied to determine $g^{(2)}(\tau)$ as defined in Eq.~\ref{eq:g2}.

\subsection{Offset measurement and normalization}

Usually, division by the mean value of $G^{(2)}$ in a time range away from the signal region ($\tau \gg \tau_\textnormal{c}$), is an appropriate way of normalizing. However, possible constant and rate-independent  offsets in the waveform baseline in one PMT channel add to $G^{(2)}$, as they are correlated with photons and offset in the other channel, and affect the normalization. While it is easy to get rid of such baseline offsets in the laboratory (e.g. by adjusting the corresponding amplifier), it cannot be easily accounted for on site at the telescopes, where accessibility of the setup is often not possible during measurements.
Constant baseline offset effects can be computed and corrected with an additional short zero-rate offset measurement, where offsets $s_A$ and $s_B$ are determined and subtracted.
If the channel waveforms $x(t) = \gamma_x(t) + s_x$ consist of photo current $\gamma_x$ and artificial offset current $s_x$, the outcome of equation \ref{eq:G2} for any time difference will on average be

\begin{align}\label{eq:offset_comp}
    \overline{G^{(2)}} &= \frac{T}{\Delta t} \cdot (\gamma_A \gamma_B + \gamma_A s_B + \gamma_B s_A + s_A s_B) \\
    &= \frac{T}{\Delta t} \cdot (\gamma_A \gamma_B) +  \frac{T}{\Delta t} (As_B + Bs_A - s_As_B)
\end{align}
where $\gamma_x$ is the mean value of $\gamma_x(t)$ within measurement time $T$, and $\Delta t$ is the sampling time bin width.
The first term on the right hand side describes the correlation of photons and the second one the contributions of offset correlations with photons and offset-offset correlations. Since only the first term is desired, the value of the second term is subtracted from every $G^{(2)}$ bin value. Afterwards, normalization by dividing every reduced $G^{(2)}$ value by the reduced $G^{(2)}$ mean (calculated outside of correlation region) is performed to obtain the $g^{(2)}$ function.

\begin{equation}
    g^{(2)} (\tau) = \frac{G^{(2)} (\tau)}{\overline{G^{(2)}}|_{\tau \gg \tau_c}}
\end{equation}
We recognize an 8-bit periodic oscillation pattern in the resulting correlation originating from the digitizer card. To correct for that, the pattern template is extracted from the $g^{(2)}$ baseline averaged over multiple 8-bit cycles and afterwards removed from every correlation bin.

\subsection{Calibration measurement}\label{sec:calibration}

Current correlation deals with photons in a different way compared to photon time tagging measurements, where the time stamp of arriving photons is stored instead of waveforms. The fact that each photon pulse seen in Fig.~\ref{fig:waveforms} extends over more than $30$ of the $1.6\,$ns time bins affects the $g^{(2)}$ function. It causes correlation of neighbouring $g^{(2)}$ bins, decreased baseline RMS expectation values (below the time tagging shot noise level) and generates a photon-pulse specific shape of the bunching peak. To quantitatively determine these effects, a short (few seconds) calibration measurement is performed before every correlation measurement by powering the halogen lamp at low current to achieve low photon rate waveforms as seen in Fig.~\ref{fig:waveforms} (blue).

From that measurement the average photon pulse shape as well as the pulse height distribution of the photons is extracted. This information is used to perform waveform simulations in order to quantitatively study the correlation baseline fluctuations, as described in section \ref{sec:results}.
Further the \lq average photon charge\rq - the PMT anode charge produced by one incoming photon on average - is calculated and used for rate determination in the correlation measurements, where the photo current in both channels is measured. The determination of the photon rates is a crucial part in order to compare the fluctuations in the $g^{(2)}$ function to the statistical noise expectation level following Eq.~\ref{eq:RMS_timetagging}. When measuring stars, the rate information will also help us to check whether the setup on the telescope is optimally adjusted and it allows for consistency checks with the expected star rates.

\begin{figure}
\centering
\includegraphics[width=1.\linewidth]{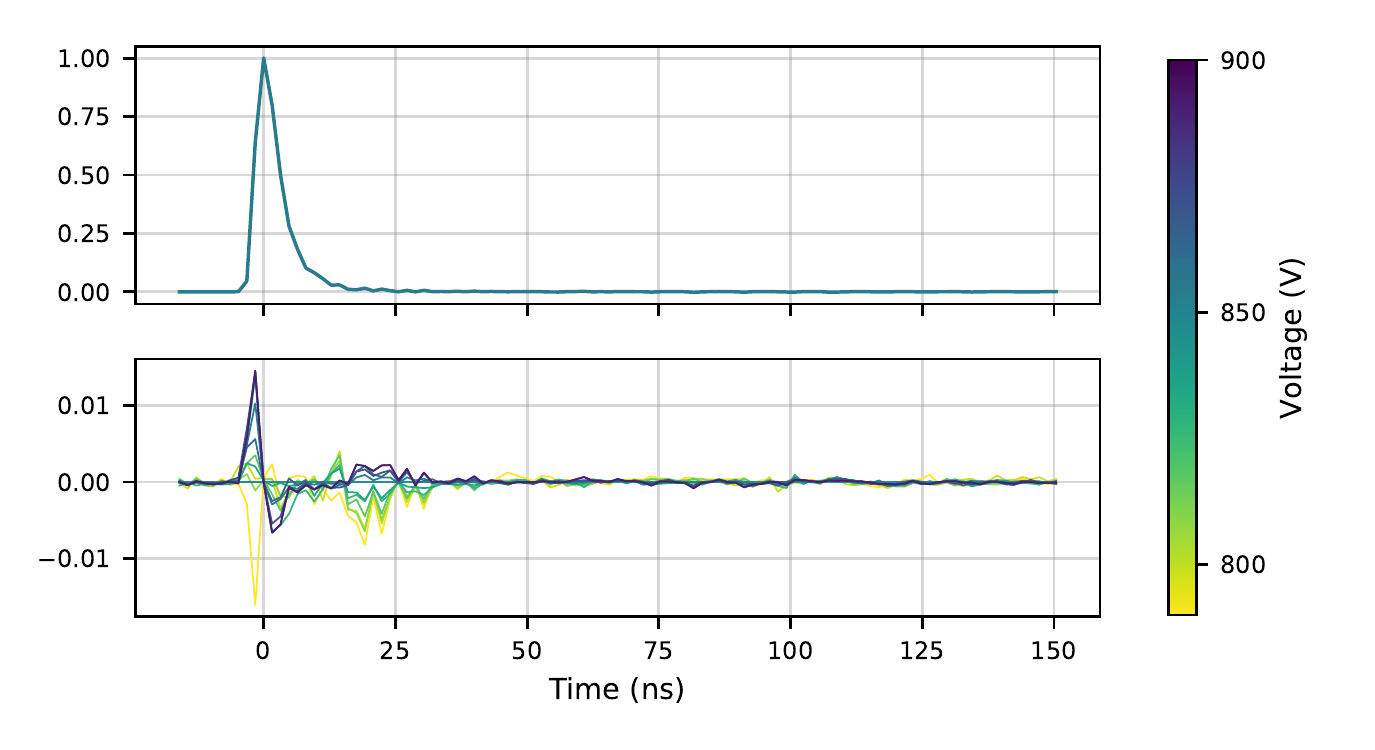}
\caption{Above: Normalized average photon pulse shape at a PMT voltage of $850\,$V. Below: the differences to upper shape at other supply voltages.}
\label{fig:cal-sub1}
\end{figure}

\begin{figure}
\centering
\includegraphics[width=1.\linewidth]{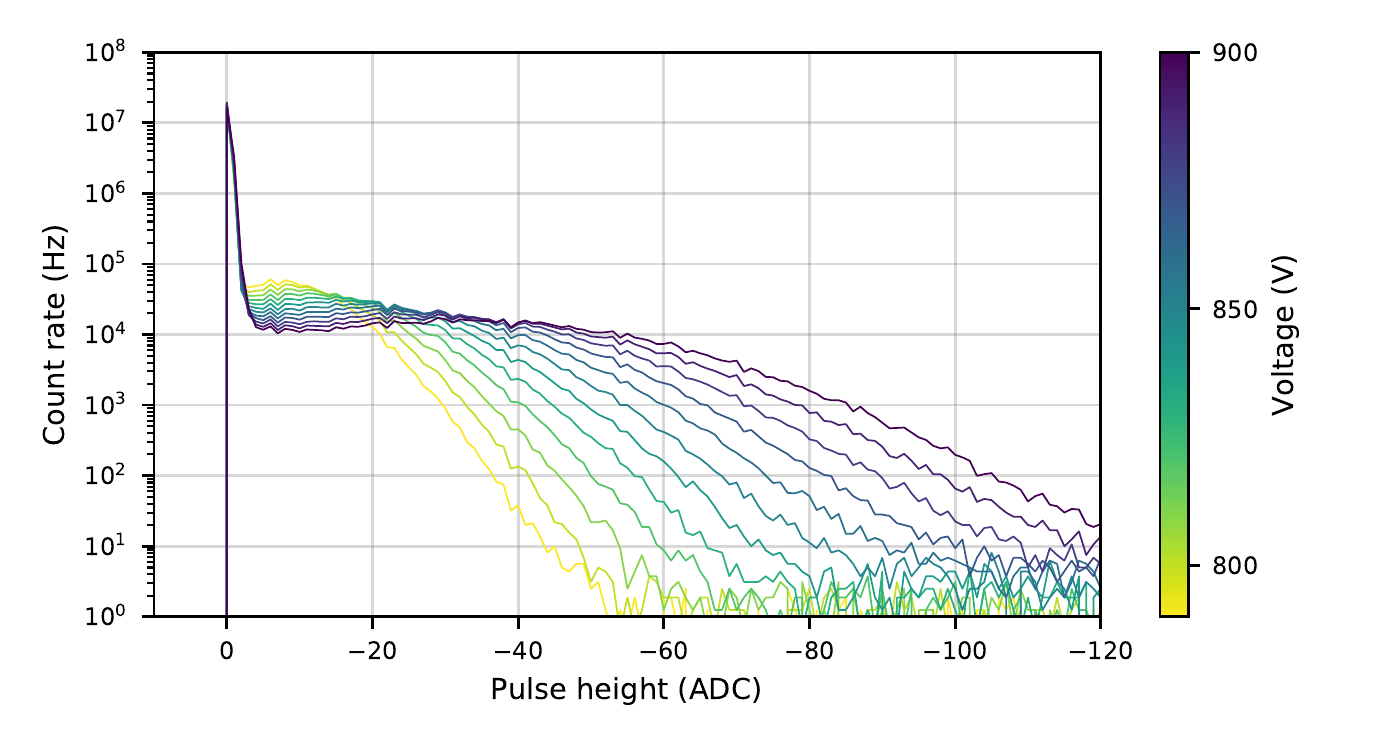}
\caption{Height distribution of single photon electron pulses at different PMT voltages. The Peak at zero is the pedestal.}
\label{fig:cal-sub2}
\end{figure}

Figs.~\ref{fig:cal-sub1} and \ref{fig:cal-sub2} show the photon pulse shapes as well as the pulse height distributions for a set of different photomultiplier voltages. While the pulse height distribution shows an expected dependency on the PMT supply voltage, the pulse shape does so only on the scale of $1$ per cent.

We use different PMT voltages in measurements at different photon rates. While at very high photon rates the voltage needs to be lowered to not exceed the maximum PMT current, it is beneficial to increase the PMT voltage and therewith gain at lower rates resulting in larger PMT pulses and clearer determination of pulse shapes and photon rates.

The pulse height distribution in Fig.~\ref{fig:cal-sub2} resembles a Gaussian distribution overlayed with a pedestal resulting from random noise spikes in the waveform baseline. A Gaussian fit is applied to the region of the curve which doesn't include the pedestal, in order to model the distribution. Since positive pulse heights are unphysical, we cut the distribution at zero in our model.

Differences in the pulse shape or the pulse height distribution appear at different environmental parameters. We use different PMTs (of same kind) and cable lengths from the photo detection to the digitizing card, which both affect the parameters. We also recognized slight changes in the pulse height distribution at different temperatures. Hence it is reasonable not to rely on pre-calibrated data, but instead carry out a calibration measurement shortly before each correlation measurement.

\section{Measurement results}\label{sec:results}
\begin{figure}
\centering
\includegraphics[width=0.9\linewidth]{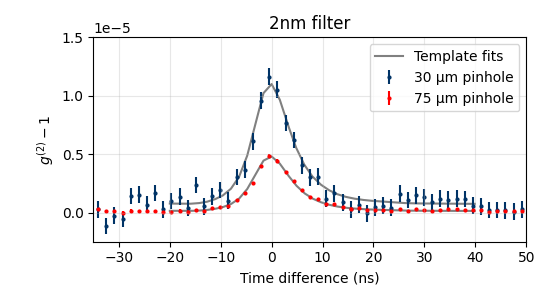}
\caption{$g^{(2)}$ functions of $1\,$hr measurement time for each pinhole with the $2\,$nm FWHM filter}
\label{fig:g2_2nm}
\end{figure}

\begin{figure}
\centering
\includegraphics[width=0.9\linewidth]{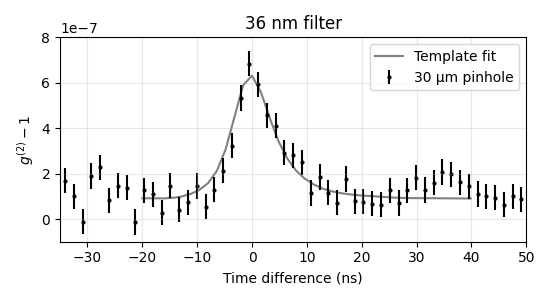}
\caption{$g^{(2)}$ function of $30\,$minutes measurement time with the $36\,$nm FWHM filter.}
\label{fig:g2_36nm}
\end{figure}

Three measurements were performed in total. We measured two different spatial coherences while using the $2\,$nm FWHM optical filter ($30$ and $75\,$\textmu m exit pinholes at the xenon lamp corresponding to $80$ and $36$ per cent spatial coherence). In each of these measurements $1\,$hr of data were recorded. We also measured at $80$ per cent spatial coherence with the $36\,$nm filter for a total of $30\,$minutes. The resulting $g^{(2)}$ functions (subtracted by $1$) are shown in Fig.~\ref{fig:g2_2nm} and \ref{fig:g2_36nm}.

We operated the two PMTs at main voltages of $820$ and $850\,$V, respectively, in order to adjust them for similar gain. For each measurement the position of the exit pinhole w.r.t. the xenon lamp and the interferometer was precisely adjusted in order to maximize the photon flux from the lamp. In case of the $2\,$nm/$30\,$\textmu m measurement, photon rates of around $270\,$MHz in one and $300\,$MHz in the other channel were recorded. This is also the order of magnitude we expect from bright stars at medium size Cherenkov telescopes for a filter width of $2\,$nm.

\begin{figure}
\centering
\includegraphics[width=0.8\linewidth]{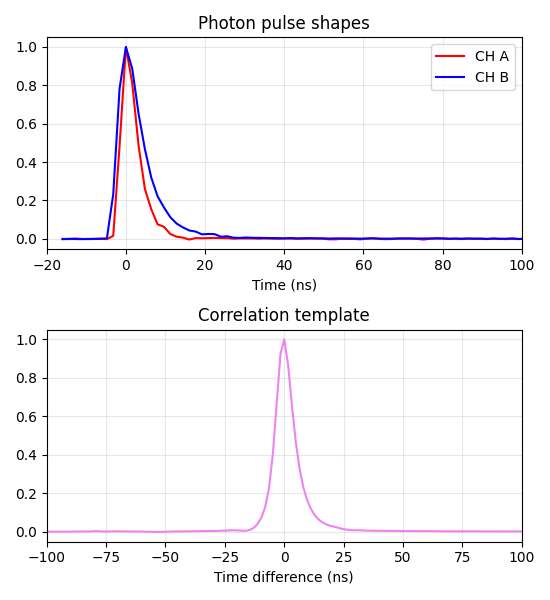}
\caption{Average single photon pulses in both channels and the resulting correlation peak shape}
\label{fig:template}
\end{figure}

A calibration measurement of $3.4\,$s duration was performed prior to the correlation measurement. Fig.~\ref{fig:template} top panel shows the average single photon pulse shapes in both channels. To avoid crosstalk in the correlation peak region, the cable between PMT/amplifier and digitizer for channel B is $30\,$m longer than the one for channel $A$. This is why its pulse shape is spread out more than the one in channel $A$. The pulse shapes are numerically correlated with each other in order to obtain a correlation peak template (Fig.~\ref{fig:template} bottom panel), which we fit to the measured $g^{(2)}$ function. The difference between the photon pulse shapes of channel $A$ and $B$ make the peak template look slightly asymmetrical.

For the other two measurements ($2\,$nm FWHM/$75\,$\textmu m pinhole and $36\,$nm FWHM/$30\,$\textmu m pinhole), the photon rates were so high that we had to decrease the PMT supply voltages to $690$ and $680\,$V in order not to damage them. As a result, the single photon pulses in the waveforms are too small to significantly distinguish them from noise spikes, which makes former calibration procedure impossible. Nevertheless, as indicated by Fig.~\ref{fig:cal-sub1} we assume the pulse shapes do not change significantly by decreasing the high voltage, nor does the peak template. Hence, the derived peak template from the lower rate PMT settings (Fig.~\ref{fig:template}) is used to fit all 3 correlation peaks in Figs.~\ref{fig:g2_2nm} and \ref{fig:g2_36nm}.

To understand whether quantitatively the correlation signal measured in the $g^{(2)}$ function matches expectations we determine the product of the coherence time $\tau_c$ - as defined in Eq.~\ref{eq:coh_time} - in combination with the spatial coherence $s$ by integration of the fitted curve, and compare it to the theoretical expectation derived numerically using the optical filter functions and the spatial coherence values obtained from the geometry ($0.8$ and $0.36$). Tab.~\ref{table:coherence_times} shows excellent agreement between measurement and expectation for all cases. Systematic uncertainties due to the choice of fit ranges may add on the order of $40-50\,$per cent of the given statistical uncertainty.

\begin{table}
\centering
\begin{tabular}{||c c c | c c||} 
 \hline
 Filter & Pinhole & Photon rates & Expected $s\cdot\tau_c$ & Measured $s\cdot\tau_c$ \\ [0.5ex] 
 \hline\hline 
 $2\,$nm  & $30\,$\textmu m & $270-300\,$MHz & $111.2\,$fs & $(112.7 \pm 4.3)\,$fs \\ 
 $2\,$nm  & $75\,$\textmu m & $2-2.3\,$GHz & $50.0\,$fs  & $(51.2 \pm 0.6)\,$fs \\
 $36\,$nm & $30\,$\textmu m & $4.3-4.7\,$GHz & $5.6\,$fs   & $(5.9 \pm 0.4)\,$fs \\ [1ex] 
 \hline
\end{tabular}
\caption{Summary of measured coherence. $s$ is the spatial coherence factor ($0.8$ for the $30\,$\textmu m pinhole and $0.36$ for the $75\,$\textmu m pinhole.)}
\label{table:coherence_times}
\end{table}

The results show that the interferometer is able to detect different signal heights due to different spatial coherence values. It proves that the system is able to measure the drop off in spatial coherence when being operated at spatially separated telescopes, which is essential for determination of the desired star properties.

Having validated the properties of the correlation signal, we turn to a quantitative understanding of the background, wich is the RMS noise of the correlation. This quantity is used to check whether there are systematic noise contributions to the correlation or whether the system is limited by photon shot-noise only. In the latter case, the RMS behaves as given in Eq.~\ref{eq:RMS_timetagging}, with the special addition of a constant factor to the equation for current correlation analysis. The  photon pulse shapes in the waveforms extend over $>30\,$ns and act as a low-pass filter to the $g^{(2)}$ function reducing the measured RMS. In the cases where the photon rates are low enough that the calibration procedure described above is possible, this factor is derived through waveform simulations, using the measured pulse shape and pulse height distribution of the photons. This method can be compared to the experimental way of determining this factor: assuming that the system works on shot noise level for short measurement times (here $T\approx\,3.4\,$s) and calculating the average RMS factor out of all $1048$ subdivisions of the measurement. These two methods are in agreement at the $7\%$ level, so that the experimental correction factor for the RMS theory curve is also used at the high-rate measurements, where single pulse calibration fails.

\begin{figure}
\centering
\includegraphics[width=0.9\linewidth]{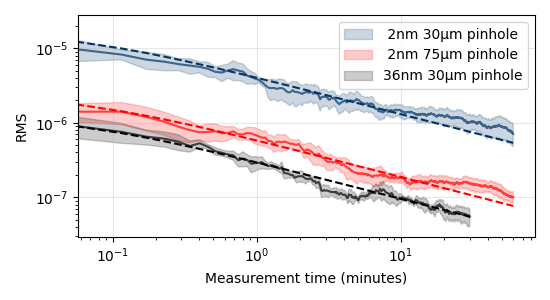}
\caption{Measured RMS noise of $g^{(2)}(\tau)$ for $\tau\gg\tau_c$ for each measurement. The dashed lines are the Shot noise expectation lines.}
\label{fig:rms}
\end{figure}

Fig.~\ref{fig:rms} shows how the RMS values of the measurements evolve with cumulative statistics. The error band is derived by calculating the $g^{(2)}$ baseline RMS in multiple sub-datasets. The dashed lines display the RMS expectation value following Eq.~\ref{eq:RMS_timetagging} including the experimental correction factor and show good agreement for all curves indicating that the measurements do not show any systematic effects on the obtained level of statistics.

However, it has to be mentioned that for the black curve this is only true if the total observed $g^{(2)}(\tau)$ is not chosen longer than $\tau = 300\,$ns, since we observed a slight, very low-frequent drop in the $g^{(2)}$ baseline, which likely is connected to temporal variations of the xenon lamp power supply.


\section{Conclusion and outlook}
We have developed an intensity inteferometry system that is capable of measuring the photon correlations of thermal light with signal-to-background values of $10^{-5}$ to $10^{-7}$ and noise level of $< 10^{-7}$ at photon rates of several hundreds of MHz up to several GHz. This enables correlation measurements of starlight with high signal to noise ratios over differen telescope baselines, where the spatial coherence has a large range of values, which  minimizes uncertainties on the measured quantities, e.g. the angular diameter of a star.
We achieved a good understanding of the current correlation system to measure the spatial coherence at different pinhole sizes (i.e. different telescope baselines) not only relative, but on an absolute scale.
We plan to operate the interferometer in a measurement campaign at Cherenkov telescopes using the beam splitter arrangement in at least two telescopes, to simultaneously measure the zero-baseline coherence as well as the reduced spatial coherence due to the telescope separation.

\section*{Acknowledgements}
This work was supported with a grant by the Deutsche Forschungsgemeinschaft (\lq Optical intensity interferometry with the H.E.S.S. gamma-ray telescopes\rq - FU 1093/3-1).

We thank Tobias Utikal and Stephan Götzinger from the Max Planck Institute for the Science of Light for the high resolution measurements of the optical filter. We thank Joachim von Zanthier, Stefan Richter and Sebastian Karl for helpful discussions and their collaboration.

\section*{Data Availability}
The data underlying this article will be shared on reasonable request to the corresponding author. Correlation histograms are available in time-intervals of \SI{3.436}{\s}. Due to the large size of the digitized waveforms in excess of several TB, the raw data can not be made available online.



\bibliographystyle{mnras}
\bibliography{cites} 








\bsp	
\label{lastpage}
\end{document}